\documentclass[tightenlines, twocolumn, notitlepage]{revtex4-1}

\usepackage{amsmath}
\usepackage{amssymb}
\usepackage{array}
\usepackage{bibunits}
\usepackage{booktabs}
\usepackage[dvipsnames]{xcolor}
\usepackage{graphicx}
\usepackage{multirow}
\usepackage{natbib}
\usepackage{tabularx}
\usepackage{units}

\newcolumntype{x}[1]{>{\centering\arraybackslash\hspace{0pt}}p{#1}}

\graphicspath{{./}{Graphics/}}
\setcitestyle{super}

\AtBeginDocument{
\heavyrulewidth=.08em
\lightrulewidth=.05em
\cmidrulewidth=.03em
\belowrulesep=.65ex
\belowbottomsep=0pt
\aboverulesep=.4ex
\abovetopsep=0pt
\cmidrulesep=\doublerulesep
\cmidrulekern=.5em
\defaultaddspace=.5em
}

\begin{document}


	\title{\Large{}Machine Learning for Quantum Dynamics: Deep Learning of Excitation Energy Transfer Properties}
	\author{Florian H\"ase}
	\author{Christoph Kreisbeck}
	\email{christophkreisbeck@gmail.com}
	\author{Al\'an Aspuru-Guzik}
	\email{aspuru@chemistry.harvard.edu}
	\address{Department of Chemistry and Chemical Biology, Harvard University, Cambridge, Massachusetts, 02138, USA}
	\date{\today}

	\begin{abstract}
	  Understanding the relationship between the structure of light-harvesting systems and their excitation energy transfer properties is of fundamental importance in many applications including the development of next generation photovoltaics. Natural light harvesting in photosynthesis shows remarkable excitation energy transfer properties, which suggests that pigment-protein complexes could serve as blueprints for the design of nature inspired devices. Mechanistic insights into energy transport dynamics can be gained by leveraging numerically involved propagation schemes such as the hierarchical equations of motion (HEOM). Solving these equations, however, is computationally costly due to the adverse scaling with the number of pigments. Therefore virtual high-throughput screening, which has become a powerful tool in material discovery, is less readily applicable for the search of novel excitonic devices. We propose the use of artificial neural networks to bypass the computational limitations of established techniques for exploring the structure-dynamics relation in excitonic systems. Once trained, our neural networks reduce computational costs by several orders of magnitudes. Our predicted transfer times and transfer efficiencies exhibit similar or even higher accuracies than frequently used approximate methods such as secular Redfield theory. 
	\end{abstract}

	\maketitle

	\begin{bibunit}[unsrt]
	\section{Introduction}

	  	Studying excitation energy transport (EET) has been of great interest across different fields bridging evolutionary biology to solar cell engineering for many years. Especially natural light-harvesting has been the subject of intense research. Pigment-protein complexes exhibit remarkable transport properties which facilitate highly efficient excitation energy transfer across long distances.\cite{caffarri2009a,baker2008a,kreisbeck2016a, amarnath2016a} Thus, identifying working principles that ultimately transform into blueprints for novel nature-inspired excitonic devices is an active research frontier.\cite{scholes2017a,scholes2011a}

	  	Mechanistic studies reveal valuable insight into the microscopic details of EET. Prominent examples are given by studies probing 
	  	the impact of electronic coherence or non-trivial interactions between excitons and specific vibrational modes on transfer characteristics.\cite{blau2017a, kreisbeck2012a, chin2013a, christensson2012a, dean2016a, Romero2014a,desio2016a} However such investigations are tedious since they require sophisticated experimental setups,\cite{dean2016a, Romero2014a,desio2016a,collini2010a,engel2007a,brixner2005a} as well as computationally involved accurate simulations of open-quantum system dynamics.\cite{blau2017a,kreisbeck2012a,chin2013a,schulze2015a,hein2012a,suess2014a,nalbach2011a} Further, there are only a few fundamentally different natural light-harvesting complexes from which alone we cannot extract the relation between the structure of an excitonic system and its dynamics in full detail. 

	  	In order to relate the dynamics to the underlying structure, it is desirable to investigate a large number of artificially designed excitonic systems. This has been recently addressed in several theoretical works.\cite{Scholak2011, Mostarda2013, Baghbanzadeh2016, Baghbanzadeh2016lett} For example, analyzing perturbations on pigment geometries in the Fenna-Matthews-Olson (FMO) complex revealed that higher transport efficiencies tend to be realized by more compact structures.\cite{Knee2017} The drawback of these statistical approaches is that they need to run exciton dynamics calculations for ten thousands of randomly generated physically-plausible multi-chromophoric structures. Due to the sheer number of performed dynamics simulations, such an analysis becomes quickly computationally exhaustive, even though less sophisticated methods such as Lindblad equations are used.\cite{Knee2017}

	  	\begin{figure*}[!ht]
	  	  	\centering
	      	\includegraphics[width = 0.95\textwidth]{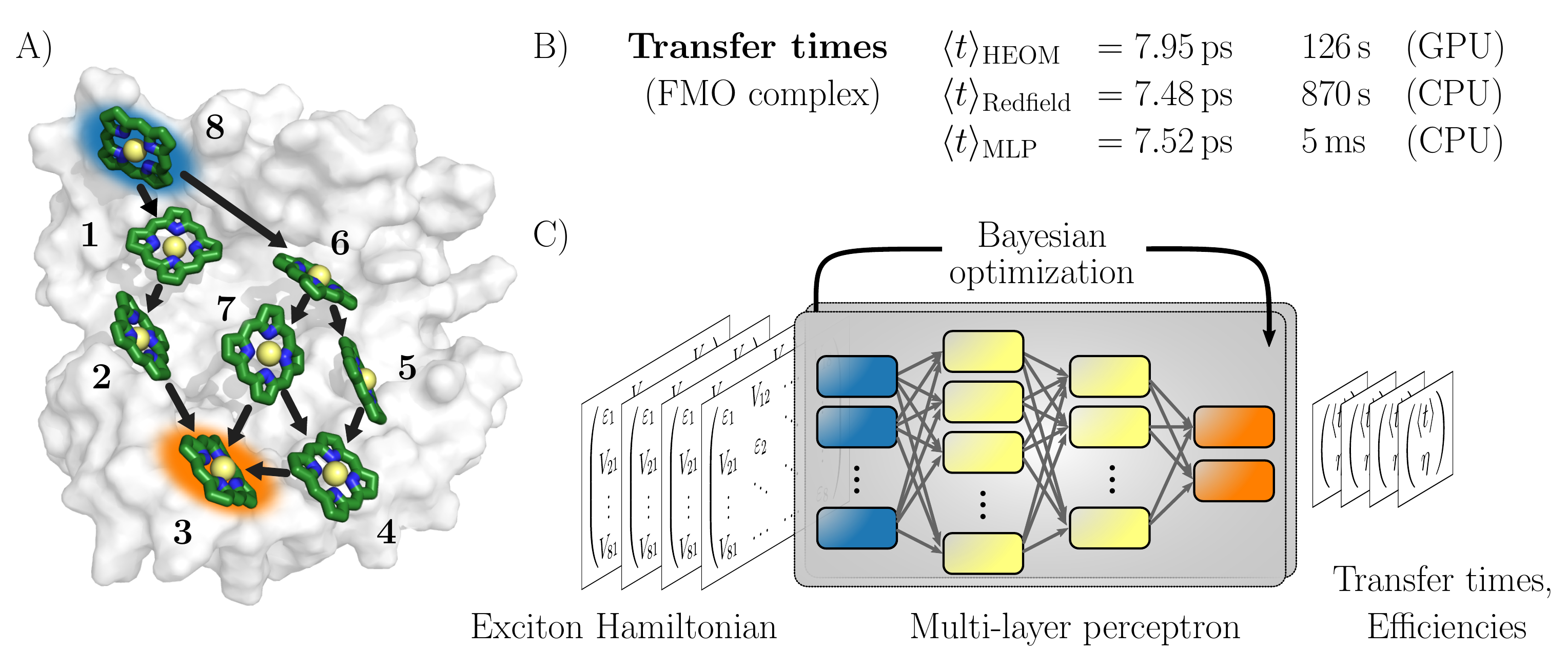}
	      	\caption{Machine learning excitation energy transfer properties in open quantum systems. (A) Fenna-Matthews-Olson (FMO) pigment-protein complex with eight chlorophyll pigments in the conventional numbering scheme. Dominant energy transfer pathways from the donor pigment~8 (blue) to the acceptor pigment~3 (orange) are indicated. (B) Results for average transfer time $\langle t \rangle$ calculations for energy transfer in the FMO complex from the donor to the acceptor obtained from solving the hierarchical equations of motion (HEOM), the approximate secular Redfield formalism and predicted by multi-layer perceptrons (MLPs) designed in this study. Computational costs are reported for each method. (C) Illustration of the MLP architecture. MLPs accept Frenkel exciton Hamiltonians as input feature and predict average transfer times and efficiencies. The best network architectures were obtained through Bayesian optimization.}
	      	\label{fig:fmo_illustration}
	 	\end{figure*}

	  	Here, we follow a novel path and leverage concepts from deep learning to bypass the computational demand of established techniques for exploring EET properties (see Fig.~\ref{fig:fmo_illustration}). Specifically, we train multi-layer perceptrons (MLPs), a class of fully connected feed-forward artificial neural networks to predict average exciton transfer times and overall transfer efficiencies. The input features to the MLPs are hereby given by the parameters of the corresponding Frenkel exciton Hamiltonians. For large scale screening of parameter space, only a fraction of all systems needs to be actually calculated to train the MLPs. 
	  	Once trained, our neural networks evaluate transfer times just within a few milliseconds and thus bypass the computational demand of established techniques for exploring EET properties, while maintaining sufficiently high prediction accuracy. 
	
	  	We demonstrate the potential of the MLPs by considering various artificial datasets which were generated by uniform sampling of pigment excitation energies and inter-pigment couplings in the vicinity of the energies and couplings of a set of relevant biological complexes: the FMO complex,\cite{Fenna1975} as well as the light-harvesting complexes CP43, CP47 and the reaction center (RC) of photosystem~II.\cite{muh2012a,raszewski2008a,raszewski2008b} We aim to predict average transfer times from an initially excited donor to a certain acceptor pigment. Fig.~(\ref{fig:fmo_illustration}) shows the situation for the FMO complex, which serves as an energy wire bridging the chlorosome and the reaction center in the photosynthetic apparatus of green sulfur bacteria and has become a standard system for comparing energy transfer properties.\cite{Mohseni2008} Initial excitation is assumed to be located at the donor pigment~8 since this pigment is in the proximity of the light-harvesting chlorosome antenna. Then, the excitation energy needs to be transferred to the target pigment~3 which couples to the reaction center where photochemical reactions are triggered. In the context of EET, the latter process is typically modeled as irreversible energy trapping.\cite{Kreisbeck2011,rebentrost2009a,caruso2009a,fassioli2010a}
	  
	  	The MLP models are trained based on transfer properties obtained with the hierarchically coupled equation of motion technique (HEOM),\cite{tanimura1989a,ishizaki2009c,tanimura2012a} which is a non-perturbative open quantum system approach taking into account non-Markovian effects. HEOM has become one of the standard tools in the field (a ready-to-run online package is available on nanohub.org)\cite{kreisbeck2013a} and serves in this manuscript as ground truth to quantify the error for the predictions made by the neural networks. The accuracy of the predictions critically depends on the choice of hyperparameters such as the number of neurons, number of hidden layers or the learning rate, which collectively define the specific architecture of the neural network. However, the best set of these parameters is \textit{a priori} unknown. Therefore, we determine the architectures for our MLP models from a Bayesian optimization on selected hyperparameters. This procedure is well-established in the machine learning community and was shown to outperform architectures built by domain experts.\cite{Snoek2012}
	  
	  	We assess the quality of our MLP predictions by comparing the relative error of our predicted transfer times to the relative error made by secular Redfield calculations. The latter is simple to implement and commonly used to avoid the numerical complexity of more accurate HEOM simulations. 
	  	Our findings demonstrate that MLPs provide a computationally significantly cheaper alternative to secular Redfield computations at comparable or, in most of our examples, even higher accuracy. Results for the FMO complex are summarized in Fig.~(\ref{fig:fmo_illustration}). 
			

    \section{Machine learning approach}\label{sec:machine_learning_approach}

		A number of studies across many fields in recent years have demonstrated how machine learning models can be utilized to accelerate a variety of computations by several orders of magnitude at a reasonable level of accuracy. For example, Gaussian processes were used to predict formation of free energies for catalyst surface chemistry.\cite{Ulissi2017} Neural networks have been successfully employed for the construction of various forms of transferable and non-transferable atomistic potentials.\cite{Behler2007, Smith2017, Yao2017} Protein-ligand binding affinities were accurately predicted by atomic convolutional neural networks,\cite{Gomes2017} and multi-layer perceptrons were trained to predict excited state energies in the context of exciton dynamics,\cite{Hase2016} as well as other electronic properties of small molecules.\cite{Hansen2013, Montavon2013}

		In the subsequent sections, we develop a machine learning framework based on multi-layer perceptrons which predict excitation energy transfer properties of excitonic systems rather than obtaining them from computationally expensive calculations. In future applications, this approach could facilitate large-scale screening such as the search for best-performing devices or studies on structure-function relationships in natural light-harvesting. 

		Overall, our procedure can be summarized as follows. Based on the Frenkel exciton Hamiltonian we leverage standard open quantum system approaches to generate a database comprising of average transfer times and efficiencies for EET from a donor to a target pigment for a random set of Frenkel exciton Hamiltonians. The complete dataset is split into a training set, on which we train each MLP model, as well as a validation and a test set. For training data selection we will compare two strategies: (i) random selection of data points and (ii) selection of training data based on a principal component analysis (PCA) which allows us to extract those data points covering the most information sampled in the dataset. As we show in Sec.~\ref{sec:results}, the latter strategy is of particular relevance if the space of transfer properties is not evenly sampled and many representatives in the training set exhibit redundant information. We run a Bayesian optimization procedure to identify the best architecture for our MLP models. The performance of each architecture is quantified by the average relative absolute error made when predicting transfer properties for the validation set. Finally, we run predictions on the test set to assess the ability of the optimized architecture to generalize to realizations that were neither employed for training nor for validation during the Bayesian optimization. The source code for exciton transfer property predictions along with all trained MLP models as well as the datasets generated in this study are made available on GitHub.\cite{githubRepo}
      
      	\subsection{Generating the excitation energy transfer database}\label{sec:dataset_generation}

		To demonstrate the capabilities of our machine learning approaches, we investigate four datasets of randomly generated excitonic systems that are sampled around pigment-protein complexes found in natural light-harvesting. For future reference, the generated database can be downloaded from a GitHub repository.\cite{githubRepo}

		For our first dataset, we sample Hamiltonians around the FMO complex (Fig.~\ref{fig:fmo_illustration}), which serves frequently as the prototype light-harvesting complex. We construct three additional datasets that are motivated by the photosystem~II of higher plants. For one set, we consider the eight pigments of the reaction center (RC) core, in which the primary step of charge separation is initiated through the electronically excited pigment Chl$_{\rm D1}$.\cite{holzwarth2006a,raszewski2008a} For the other two sets, the reaction center core is extended by including either light-harvesting complex CP47 or CP43 of photosystem~II into the exciton system. For simplicity, we refer to the dataset inspired by the CP43+RC (CP47+RC) complex as the CP43 (CP47) dataset from hereon. For each dataset, we generated 12000 exciton Hamiltonians by uniformly sampling excited state energies and inter-site couplings from a fixed range of values, as is summarized in Tab.~\ref{tab:hamiltonian_ranges}.  

		\begin{table}[!t]
		  \centering
		  \begin{tabular}{lccccc}
	    		\toprule
	    		Label & \#Sites & $\unit[\varepsilon_\text{low}]{[cm^{-1}]}$ & \quad$\unit[\varepsilon_\text{high}]{[cm^{-1}]}$  & \quad$\unit[V_\text{range}]{[cm^{-1}]}$ \\
	    		\midrule
				    RC        &  8 & 14800 & 15000 &  -50 ...  50 \\
				    FMO       &  8 & 12000 & 12800 & -100 ... 100 \\
				    CP43      & 21 & 14800 & 15100 &  -60 ...  60 \\
				    CP47      & 24 & 14500 & 15300 & -100 ... 100 \\
	    		\bottomrule
		   \end{tabular}
		   \caption{Lower and upper limits in between which excited state energies $\varepsilon$ and inter-site couplings $V$ were sampled uniformly to generate the four datasets of this study. Each dataset consists of 12000 Hamiltonians with excited state energies and inter-site couplings within the reported ranges. Note, that the labels CP43 (CP47) denote datasets which are inspired by the CP43+RC (CP47+RC) biological complexes.}
		   \label{tab:hamiltonian_ranges}
		\end{table}		

		In the following, we are interested in transfer characteristics such as average transfer times from an initially excited pigment (donor) to a target pigment (acceptor). This model provides a simple description of the first step of photosynthesis, where energy is absorbed in the antenna pigments and subsequently transferred to the reaction center in which photochemical reactions are triggered. The energy transport in light-harvesting complexes is determined by coupled pigments which are embedded in a protein scaffold,\cite{May2004a, Cheng2009} and is typically modeled with an effective Frenkel exciton Hamiltonian.\cite{Leegwater1996, May2008} We include energy trapping in the acceptor pigment phenomenologically by introducing anti-Hermitian parts in the Hamiltonian. The exciton dynamics is expressed in terms of the reduced density matrix, which can be obtained from standard open quantum system approaches. 

		We compute exciton transfer times for all Hamiltonians in our datasets with the hierarchical equations of motion (HEOM)\cite{tanimura1989a,ishizaki2009c,tanimura2012a} method, implemented in the \emph{QMaster} software package, version 0.2.\cite{Kreisbeck2011, Kreisbeck2012, Kreisbeck2014} HEOM is a numerically exact method which accurately accounts for the reorganization process,\cite{Yan2004, Xu2005, Ishizaki2005, Ishizaki2009} in which the vibrational coordinates rearrange to their new equilibrium positions upon electronic transition from the ground to the excited potential energy surface. For all Hamiltonians we assumed identical Drude-Lorentz spectral densities $J(\omega)=2\lambda \frac{\omega \nu}{\omega^2+\nu^2}$, describing the exciton-phonon interaction. We do not use the parameters of the spectral density as input features for our neural networks. Extending our approach to predict transfer properties for various spectral densities goes beyond the present scope and is the aim of future work. More details on the Frenkel exciton Hamiltonian and the exciton dynamics methods, as well as the definition of the transfer time and transfer efficiencies, are given in the supplementary information Sec.~\ref{sec:modeling_exciton_transfer}. 
		\begin{figure}[!t]
		      \centering
		      \includegraphics[width = 0.98\columnwidth]{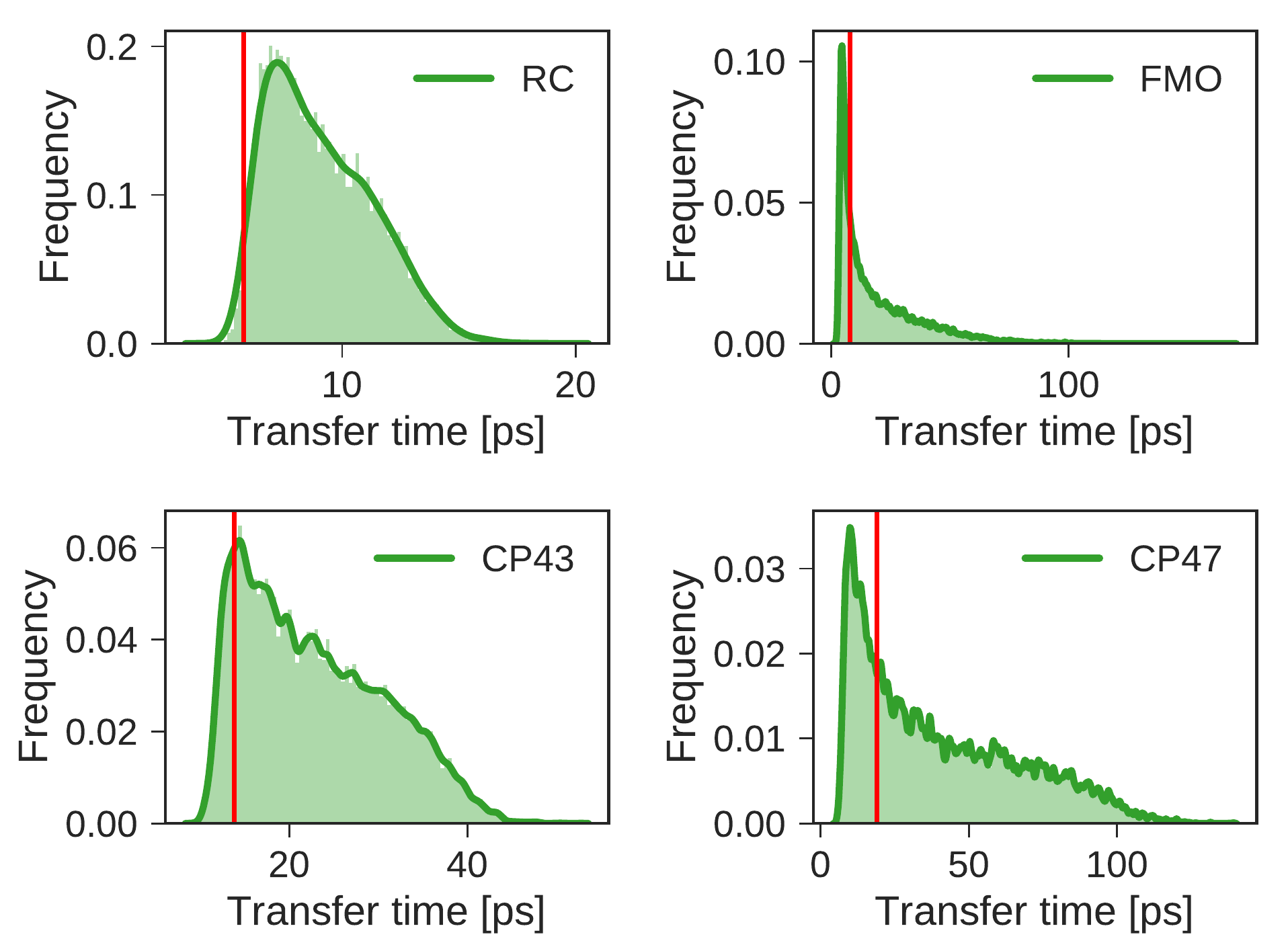}
		      \caption{Distributions of exciton transfer times computed for all 12000 generated exciton Hamiltonians for each dataset using the HEOM approach implemented in \textit{QMaster}. Vertical red lines indicate the transfer time of the exciton Hamiltonian corresponding to the biological complex. In all calculations we use a trapping rate of $\Gamma_\text{trap}^{-1}=1$~ps, an exciton life-times of $\Gamma_\text{loss}^{-1}=0.25$~ns, and a temperature of $T=300$~K. The parameters of the spectral density are set to $\lambda = \unit[35]{cm^{-1}}$, $\nu^{-1} = \unit[50]{fs}$.}
		      \label{fig:transfer_time_distributions}
		\end{figure}
	
		Distributions of transfer times for all exciton Hamiltonians of each dataset are depicted in Fig.~\ref{fig:transfer_time_distributions}. The transfer times for the Hamiltonians of the biological complexes are highlighted in every distribution. Excited states and inter-site couplings for the exciton Hamiltonians of the biological complexes are taken from literature,\cite{adolphs2006a,muh2012a,raszewski2008a,raszewski2008b} and are uploaded to the GitHub repository.\cite{githubRepo} All population dynamics simulations are initialized as a fully populated site~1, serving as a donor, while site~3 acts as acceptor that couples to an energy sink with trapping rate $\Gamma_\text{trap}$ (see supplementary information Sec.~\ref{sec:modeling_exciton_transfer}). Note that the labeling of the donor and acceptor state is without loss of generality as rows and columns of the Hamiltonian can be permuted in a suitable way, which effectively corresponds to a relabeling of the pigments. 

		We find large variations in the ranges of transfer times between the four datasets. The RC and CP43 datasets, both with relatively narrow ranges of excited state energies and site couplings, yield relatively small transfer times. In contrast, we observe a wider spread in transfer times for the FMO dataset and the CP47 dataset which is consistent with the broader range of excited state energies and site couplings that were sampled. 

		The transfer times of the actual biological complexes lie close to the mode of the distributions for all four datasets. This suggests that natural systems may not be specifically selected for extraordinary transfer properties, as they exhibit transport characteristics that are just likely to occur, even for a random choice of the exciton Hamiltonian. The conclusions of a recent evolutionary study for the FMO complex,\cite{Valleau2017} goes along a similar direction and suggests that the FMO complex has evolved towards stability to mutations rather than a selection of specific transfer characteristics. However, we note that we did not take into account structural considerations which could change the picture as many of our randomly generated artificial Hamiltonians may not be realizable under structural constraints.

      	\subsection{Principal component analysis for improved training data selection}\label{sec:training_data_selection}

		We select the training sets for our MLP models following two methods for dataset splitting. In the simplest ansatz, we select the training set randomly from our created dataset. However, due to the nature of how we randomly sampled our Hamiltonians, the transfer characteristics are not distributed homogeneously and many representations of our Hamiltonians might be very similar and thus are expected to carry redundant information. As can be seen in Fig.~\ref{fig:transfer_time_distributions}, Hamiltonians yielding longer transfer time-scales are for example underrepresented in all four datasets. 

		Therefore, we follow a different path and carry out a more sophisticated selection process. The idea is to add those Hamiltonians to our training set which give the most information. We perform a principal component analysis (PCA) on the 8000 Hamiltonians containing dataset (after separating 2000 Hamiltonians each for validation and testing). We project each Hamiltonian onto a reduced space spanned by the most relevant principal components. The Hamiltonians for the training set are then selected such that they are maximally separated in the reduced space. This procedure guarantees that our training set constitutes the most diverse entities.

	\subsection{Setup of the multi-layer perceptron architecture}\label{sec:machine_learning_aspects}
	
	  	The architectures of our multi-layer perceptrons (MLPs) are designed for supervised learning of exciton energy transfer properties. All exciton Hamiltonians were reshaped into vectors and provided as input features to the MLPs, which were used to predict exciton transfer times and transfer efficiencies simultaneously. Since, the input features of neural networks need to be of fixed size, we construct separate MLPs for each dataset in order to treat the different dimensionalities of the exciton Hamiltonians. Details on the rescaling of the input features and predicted output, as well as on the training procedure are provided in the supplementary information (see Sec.~\ref{sec:mlp_comments}).
	  
		The 12000 Hamiltonians of each dataset were split into three sets: a training set of up to 6000 Hamiltonians for training MLP model instances with particular hyperparameters, a validation set of 2000 Hamiltonians used to evaluate the MLP architecture during optimization of the hyperparameters and a test set of 2000 Hamiltonians to probe out-of-sample prediction accuracies. All constructed MLP models were trained with stochastic gradient descent with 200 data points per batch and the ADAM optimizer,\cite{Kingma2014} until the average relative absolute error (see Eq.~\ref{eq:relative_time_deviation}) on the validation set increased over three full consecutive training epochs. Neuron saturation was avoided with L2 regularization on all weights of all neurons but the output neurons. 
	  
	  	An essential component in developing accurate machine learning models consists in choosing proper values for the model hyperparameters. For this MLP framework, we consider a total of six hyperparameters. The initial learning rate $\mu$ for the ADAM optimizer and the regularization parameter $\lambda$. We also included the number of MLP layers and the number of neurons per layer, as well as the activation functions for neurons in each layer, for which we allowed five different options to choose from. The only exception is the last layer, for which we always use the softplus activation function to constrain our MLP models to the prediction of always positive transfer times and efficiencies. Lastly, we treat the number of training points as a hyperparameter in order to study the effect of the variations in the number of training samples on the prediction accuracy. The set of hyperparameters to be optimized and their allowed ranges are summarized in the supplementary information in Tab.~\ref{tab:bayes_opt_hyperparameter_selection} 

	  	We employ a Bayesian optimization algorithm,\cite{Dixon1978} in order to scan the space of hyperparameters for the most accurate model. The model accuracy was defined as the average relative absolute error (see Eq.~\ref{eq:relative_time_deviation}) in exciton transfer times predicted by the MLP and corresponding HEOM simulations for the validation set. All generated MLP models were constructed and trained with the same random seed. Bayesian optimization is a common tool in machine learning and balances exploration of parameter space and exploitation of previous information. The idea of this ansatz is to reduce the number of costly function evaluations under the assumption that the unknown function was sampled from a Gaussian process. In contrast to gradient or Hessian based optimization techniques, Bayesian optimization uses information of all previously evaluated points and can thus find a good approximation to the minimum of non-convex functions in relatively few iterations. We carried out the Bayesian optimization of MLP hyperparameters in the spearmint software package.\cite{Snoek2012} MLP models were generated and trained using the Tensorflow package, version 1.0.\cite{tensorflow2015-whitepaper}
	
		\begin{table}[!t]
			\centering
			\begin{tabular}{llccc}
				\toprule
				Dataset    			  &  Model 		   & \unit[$\Delta\tau_\text{train}$]{[\%]} & \unit[$\Delta\tau_\text{valid}$]{[\%]} & \unit[$\Delta\tau_\text{test}$]{[\%]} \\	
				\midrule
				\multirow{2}{*}{FMO}  &  Network (PCA) &  \textbf{4.53} 	&  \textbf{4.38}	 &   \textbf{7.41}   \\
						      &  Network       & 10.53  & 10.75  &  11.56   \\
						      &  Redfield      &  9.70  &  9.96  &   9.60	\\                 
				\midrule
				\multirow{2}{*}{RC}   &  Network (PCA) &  \textbf{2.71}	&  \textbf{2.73}	 &   \textbf{3.35} 	\\
						      &  Network       &  3.61  &  3.58  &   3.76   \\
						      &  Redfield      &  8.62 	&  8.67	 &   8.60 	\\
				\midrule
				\multirow{2}{*}{CP43} &  Network (PCA) &  \textbf{4.42}	&  \textbf{4.47}  &   \textbf{4.72}	\\
						      &  Network       &  4.66  &  4.71  &   4.86   \\
						      &  Redfield   &  4.71  &  4.66  &   4.73	\\
				\midrule
				\multirow{2}{*}{CP47} &  Network (PCA) & 12.36	& 12.32	 &  12.59  	\\
						      &  Network       & 13.36  & 13.34  &  13.59   \\
						      &  Redfield   & \textbf{10.48}  & \textbf{10.47}  &  \textbf{10.51}	\\
				\bottomrule
			\end{tabular}
			\caption{Average relative absolute error $\Delta\tau$ (see Eq.~\ref{eq:relative_time_deviation}) of exciton transfer times computed with HEOM and either, predicted by the trained neural networks (with/without PCA selection) or computed with secular Redfield. For all four datasets, we show the results of the training, validation, and test set separately. Smallest errors for each dataset are printed in bold.}
			\label{tab:mad_redfield_networks}
		\end{table}

	\section{Results: Prediction of transfer times with neural networks}\label{sec:results}
	
	  	In the subsequent discussion, we demonstrate the capabilities of our trained MLP models by analyzing the average relative absolute error
		  	\begin{equation}\label{eq:relative_time_deviation}
			      \Delta\tau= \Big\langle \frac{| t_\text{HEOM} - t_\text{model}|}{t_\text{HEOM}} \Big\rangle_\text{dataset},
	  	\end{equation}
	  	between predicted exciton transfer times and the ones obtained with the numerically exact HEOM calculations. Although we restrict our discussion to transfer times, we note that similar conclusions hold for the analysis of the transfer efficiencies since both characteristics are strongly correlated. Tab.~\ref{tab:mad_redfield_networks} summarizes the results for the predicted transfer times for our four generated datasets.  
	
	  	The predictions are carried out with the Bayesian optimized MLP architectures, which show slight variations in their best-performing hyperparameters depending on the dataset at hand. However, for all datasets, the neural networks tend to prefer shallow but broad architectures comprising of only a few layers with each layer containing a larger number of neurons. More details on the procedure and results for the hyperparameter optimization can be found in the supplementary information Sec.~\ref{sec:sup_bayesian_optimization}.
	
	\subsection{Prediction accuracies of trained multi-layer perceptrons}

\begin{figure}[!t]
			\centering
			\includegraphics[width = 0.92\columnwidth]{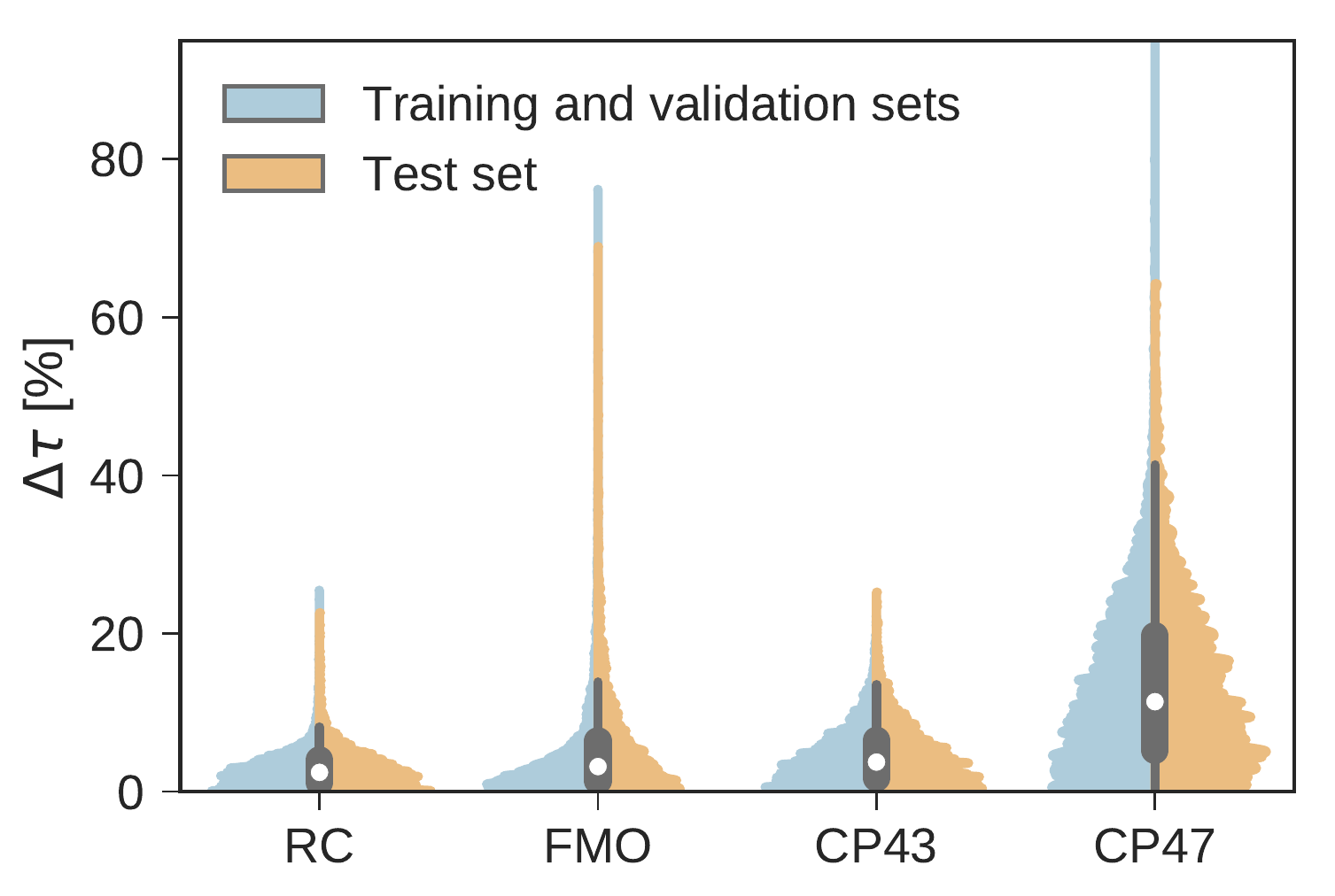}
			\caption{Normalized distributions of the average relative absolute error of predicted exciton transfer times and exciton transfer times computed with HEOM. The left (blue) side of the plots illustrate the distributions of average relative absolute errors for predictions on the training and the validation set, while the right (orange) side of the plots illustrates the errors for predictions on the test set.}
			\label{fig:out_of_sample_predictions}
		\end{figure}

		Our trained MLP models predict exciton transfer times for out-of-sample Hamiltonians at almost the same accuracy as for Hamiltonians on which MLP parameters and hyperparameters were optimized (see Tab.~\ref{tab:mad_redfield_networks}). This demonstrates the ability of our MLP models to generalize to previously unseen data and to provide accurate out-of-sample predictions. Noteworthy, there is no significant asymmetry in the distribution of the relative absolute errors for the individual Hamiltonians or the training/validation and test set (see Fig.~\ref{fig:out_of_sample_predictions}). Therefore, the architectures of the neural networks are well-balanced and neither in the regime of over-fitting, which would result in a large discrepancy in errors between the training and validation sets nor did we over-optimize the neural network architecture during Bayesian optimization. 
		
		Overall we find a high accuracy of our predictions and small average relative errors on the test sets which are in the range between \unit[3.35]{\%} for RC (PCA selected training set) and \unit[13.59]{\%} for the largest considered exciton system CP47 attached to RC (random selected training set). The CP47 dataset exhibits the most diverse transfer properties (see Fig.~\ref{fig:transfer_time_distributions}), which explains the larger average relative absolute errors in the predictions when compared to the other datasets. 
		
		The accuracy of the predictions can be enhanced by a more sophisticated PCA selection of the training set without the need of generating additional computationally expensive data points. The level of improvement of the PCA selection over a random selection of the training set differs for the four complexes. In general, we find that MLPs can be trained almost equally accurate with either selection method. The highest benefit of the PCA selected training set is obtained for the FMO and CP47 dataset, which are not only the most diverse ones out of our four datasets but are biased towards Hamiltonians showing fast transfer. As intuitively expected, selecting training points based on PCA is most advantageous for datasets with an extremely unevenly sampled feature space.

	\subsection{Comparing multi-layer perceptron predictions to secular Redfield results}
	
		Next, we provide a context for the observed MLP prediction accuracies by comparing them to the errors made by the frequently employed secular Redfield method, which is essentially derived from second order perturbation theory in the system-bath interaction in combination with a Markov approximation. Accuracies of the transfer times for both, the secular Redfield calculations and the MLP predictions are evaluated according to Eq.~(\ref{eq:relative_time_deviation}). Here, the HEOM calculations again serve as ground truth. For the datasets inspired by the smaller exciton systems FMO and RC, the trained MLPs outperform secular Redfield, even for out-of-sample predictions, whereas for the datasets around larger systems both approaches are similarly accurate. 

		For example in the case of the biological exciton Hamiltonian of the FMO complex, HEOM reveals a transfer time of \unit[7.95]{ps}. The trained MLP model predicts a transfer time of \unit[7.52]{ps} which is slightly more accurate than secular Redfield calculations that result in \unit[7.48]{ps}. Exciton transfer times obtained for all four biological complexes with all three approaches are reported in the supplementary information Tab.~\ref{tab:transfer_times_bio_complexes}. However, while the MLP prediction takes about 5~ms, secular Redfield calculations took about \unit[14.5]{min} on a single CPU (computation times are listed in the supplementary information in Tab.~\ref{tab:transfer_time_computing_times}). We conclude that our trained MLP predictions are competitive to secular Redfield calculations in terms of their accuracy, but (once trained) come at a significantly reduced computational cost.
	
		\begin{figure}[!b]
			\centering
			\includegraphics[width = 0.95\columnwidth]{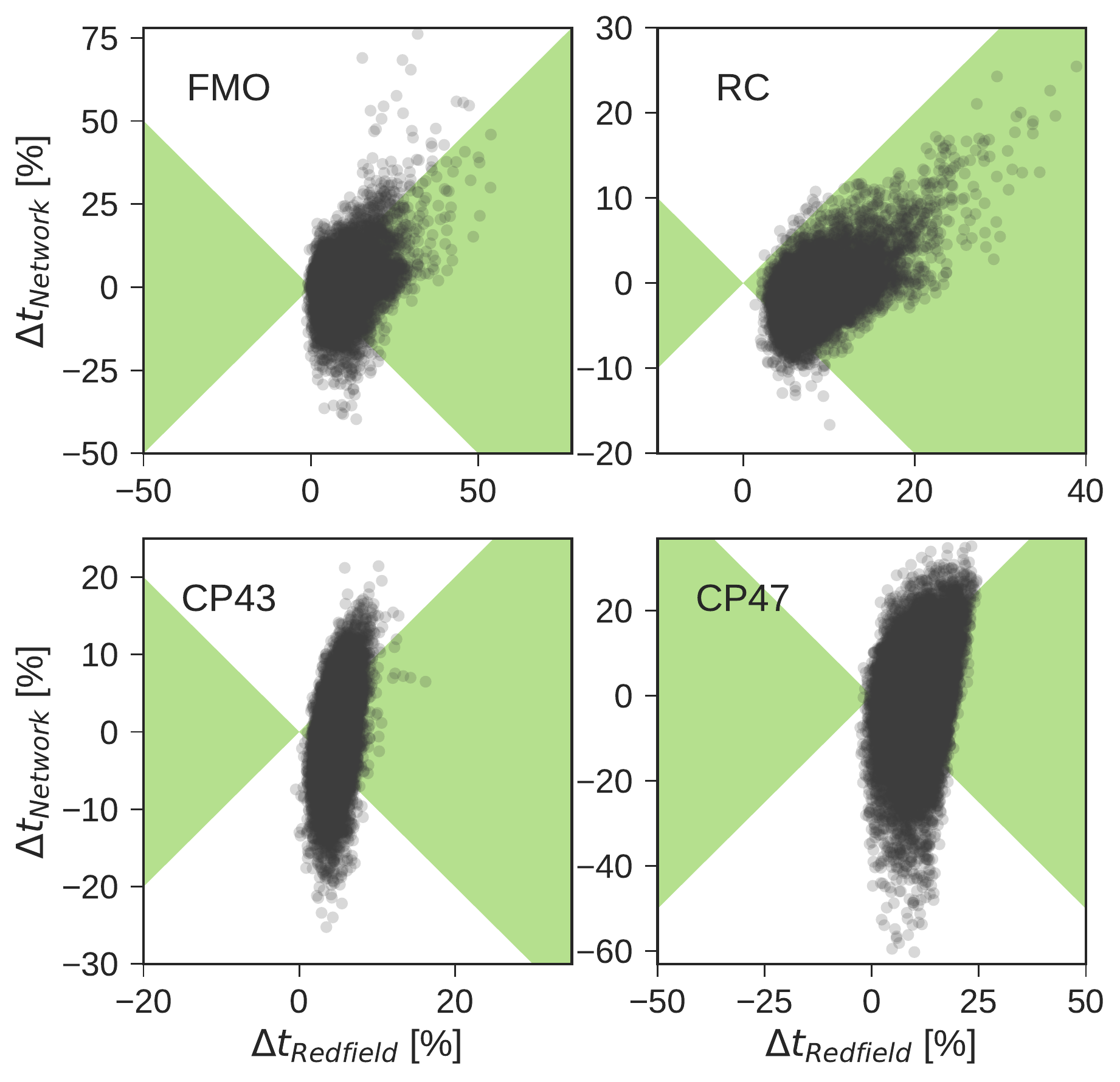}
			\caption{Relative errors in exciton transfer times computed with the hierarchical equations of motion (HEOM) approach and exciton transfer times computed with the secular Redfield approach and predicted by neural networks respectively. Displayed are relative deviations for all four datasets: the Fenna-Matthews-Olson (FMO) complex, the reaction center (RC) core, the RC with the CP43 complex and the RC with the CP47 complex. Regions in which the absolute of deviations of neural network predicted transfer times from HEOM computed transfer times are smaller than deviations for Redfield are shaded in green.}
			\label{fig:scatter_comparisons_networks_redfield}
		\end{figure}
	
		Besides analyzing the accuracy in terms of averaging over all realizations in the datasets, we compare the relative errors in transfer time for secular Redfield and the MLP predictions in more detail on the level of individual Hamiltonians. Fig.~\ref{fig:scatter_comparisons_networks_redfield} depicts scatter plots where the horizontal axes measure the accuracy of secular Redfield calculations and the vertical axes reflect the accuracy of MLP predictions for MLPs trained on the PCA selected datasets. We do not distinguish between training, validation, and test set and show the complete dataset. Almost all the Hamiltonians show a $\Delta t_\text{Redfield}=(t_\text{HEOM} - t_\text{Redfield})/t_\text{HEOM}>0$, which demonstrates that secular Redfield systematically underestimates transfer time scales. On the other hand, the predictions under- as well as overestimate transfer time-scales yielding a more symmetrical distribution along the horizontal axis. For the RC (FMO) dataset, more than \unit[95]{\%} (\unit[80]{\%}) of the Hamiltonians fall into regions marked as green, for which the neural networks provide higher accuracy than secular Redfield. For all other datasets, secular Redfield and the MLP predictions are equally likely to give better results, with about \unit[59]{\%} (\unit[57]{\%}) of the Hamiltonians for CP43 (CP47) falling within the green shaded region. This is in agreement with our average relative absolute errors listed in Tab.~\ref{tab:mad_redfield_networks}.  We did not observe any cases for which the MLPs show relative errors that significantly exceeded any of the secular Redfield ones. \\



	
	\section*{Conclusion}
	
	In this study, we have outlined how machine learning approaches can be employed to bypass computationally costly simulations of open quantum system dynamics in the context of excitation energy transfer. Overall we find that MLPs are capable of predicting transfer times for excitonic systems at higher or comparable accuracy than the frequently used secular Redfield approach albeit at much lower computational costs. Therefore we conclude that MLP models are a promising alternative for extracting excitation energy transfer properties when compared to frequently used rate equation methods.
	
	The presented approach is of particular interest for large-scale analyses of the structure-transport relationship in excitonic systems. An area of great interest in excitonics is the study of the dynamics of charge dissociation at the interface present in bulk heterojunction photovoltaics.\cite{Jailaubekov2012, Vithanage2013} We believe a tool like this will help in the rapid screening of material properties in the mesoscale and therefore help the search for high-performance OPV systems.\cite{Hachmann2011} 
	
	Once trained, evaluations of MLP models come at almost no additional cost. Our four generated MLP architectures (each optimized for one of the four datasets) predict transfer times for an aggregated set of 48,000 exciton Hamiltonians just within a few seconds, while the corresponding quantum dynamics simulations take several GPU (CPU) years for the HEOM (secular Redfield) calculations. Our trained MLP models extend well to out-of-sample predictions for exciton Hamiltonians that are close to the sampled parameter regime. However, to employ MLPs on parameter regimes beyond those probed in the existing database requires running computationally expensive exciton dynamics for a few thousand Hamiltonians in order to extend our training set. To avoid this bottleneck a potential strategy could be to leverage already existing data, e.g. produced by a user community of existing software packages such as \textit{QMaster}. However such data can be quite diverse. To this end, future research needs to focus on novel more general neural network architectures that accurately predict transfer times for flexible spectral density parameters as well as for differently sized exciton systems.
	

	\section*{Acknowledgments}

	F.H. is supported by the Herchel Smith Graduate Fellowship. C.K. is supported by the National Science Foundation under award number CHE-1464862. A.A.-G. acknowledges support from the Center for Excitonics and Energy Frontier Research Center funded by the U.S. Department of Energy under award DE-SC0001088. All computations reported in this paper were completed on the Odyssey cluster supported by the FAS Division of Science, Research Computing Group at Harvard University.


	\phantomsection\addcontentsline{toc}{section}{\refname}\putbib[main]	
	\end{bibunit}

	\clearpage
	\newpage

	\begin{bibunit}[unsrt]
%
%
%
%
%
%
%
%

\onecolumngrid
\setcounter{subsection}{0}

\section*{Supplementary Information}

	\subsection{Modeling of excitation energy transfer}\label{sec:modeling_exciton_transfer}

			The energy transport in light-harvesting complexes is determined by coupled pigments which are embedded in a protein scaffold.\cite{May2004a, Cheng2009} The large number of degrees of freedom in the system renders \textit{ab initio} calculations on the atomistic level impossible. Therefore, the exciton transfer dynamics is typically modeled with an effective Frenkel exciton Hamiltonian.\cite{Leegwater1996, May2008} The exciton Hamiltonian for a system of $N$ sites for the single exciton manifold reads
			\begin{align}
	   			H_\text{system} = \sum\limits_{i = 1}^N \varepsilon_i |i\rangle\langle i| + \sum\limits_{i \neq j}^N V_{ij} |i\rangle \langle j|,
			\end{align}
			where $\varepsilon_i$ denotes the first excited state energy of the $i$-th pigment molecule and $V_{ij}$ denotes the Coulomb coupling between excited states at the $i$-th and $j$-th molecule. We assume that the exciton system couples linearly to the vibrational environment of each pigment, which is assumed to be given by a set of harmonic oscillators. The phonon mode dependent interaction strength is captured by the spectral density
			\begin{equation}\label{eq:SpecDens}
	  			J_i(\omega)=\pi\sum_k \hbar^2\omega_{i,k}^2 d_{i,k}^2\delta(\omega-\omega_{i,k}).
			\end{equation}
			Here, $d_{i,k}$ defines the coupling of the $k$-th phonon mode ($b^\dagger_{i,k}$) of the $i$-th pigment with frequency $\hbar \omega_{i,k}$. \\

			In the first step of photosynthesis, energy is absorbed in the antenna pigments and subsequently transferred to the reaction center in which photochemical reactions are triggered. Within a simple picture, this process can be modeled by energy transfer from an initially excited pigment (donor) to a target state (acceptor). We model energy trapping in the acceptor state $|\text{acceptor} \rangle$ phenomenologically by introducing anti-Hermitian parts in the Hamiltonian
			\begin{equation}
	  			\mathcal{H}_\text{trap} = -i\hbar\Gamma_{\rm trap}/2\,|\mbox{acceptor}\rangle\langle\mbox{acceptor}|,
			\end{equation}
			where $\Gamma_\text{trap}$ defines the trapping rate. In a similar way, we model radiative or non-radiative decay to the electronic ground state as exciton losses
			\begin{equation}
	  			\mathcal{H}_\text{loss} = -i\hbar\Gamma_{\rm loss}/2\,\sum_i|i\rangle\langle i|.
			\end{equation}
			The rate $\Gamma_\text{loss}^{-1}$ defines the exciton lifetime. In this study we are interested in two different exciton propagation characteristics: the average transfer time 
			\begin{align}\label{eq:transfer_time_definition}
	  			\langle t \rangle=\Gamma_\text{trap}/\eta \int_0^{t_{\rm max}} \mbox{d}t\, t \, \langle\mbox{acceptor}|\rho(t)|\mbox{acceptor}\rangle,
			\end{align}
			and the overall efficiency 
			\begin{align}\label{eq:transfer_efficiency}
	  			\eta=\int_0^{t_{max}}\mbox{d}t\, \Gamma_\text{trap}  \langle\mbox{acceptor}|\rho(t)|\mbox{acceptor}\rangle,
			\end{align}
			which corresponds to the accumulated trapped population during the transfer process. For numerical evaluations, we replace the upper integration limit by $t_{max}$ which is chosen such that the total population within the pigments has dropped below $0.0001$. \\

			The exciton dynamics is expressed in terms of the reduced density matrix $\rho(t)$, which can be obtained from standard open quantum system approaches. Here we employ the hierarchical equations of motion (HEOM) approach which accounts for the reorganization process,\cite{Yan2004, Xu2005, Ishizaki2005, Ishizaki2009} in which the vibrational coordinates rearrange to their new equilibrium position upon electronic transition from the ground to the excited potential energy surface. The major drawback of the HEOM approach is the adverse computational scaling, which arises from the need to propagate a complete hierarchy of auxiliary matrices. Therefore, we employ a high-performance implementation of HEOM integrated into the \textit{QMaster} software package.\cite{Kreisbeck2011, Kreisbeck2012, Kreisbeck2014} As demonstrated in previous publications, \textit{QMaster} enables HEOM simulations for large systems comprising of up to hundred pigments,\cite{kreisbeck2016a} as well as to perform accurate calculations for highly structured spectral densities.\cite{blau2017a, kreisbeck2013, Kreisbeck2014, Kreisbeck2012} \\

			A computationally much cheaper formalism, the Redfield approach, can be derived with the assumption of weak couplings between the system and the bath in combination with a Markov approximation.\cite{Breuer2002, May2008} The secular approximation simplifies the equation even further and allows to write the dynamics in the form of a Lindblad master equation. This drastically reduces the computational demand of this approach compared to exciton propagation under the HEOM, which explains the popularity of the secular Redfield equations. However, secular Redfield has been shown to underestimate the transfer times in certain light-harvesting complexes.\cite{Novoderezhkin2011}

	\subsection{Exciton transfer properties of biological Hamiltonians}\label{sec:bio_transfer_times}

		We report the exciton transfer times calculated for the four light-harvesting complexes (LHCs) FMO, RC, CP43+RC and CP47+RC.\cite{adolphs2006a, muh2012a, raszewski2008a, raszewski2008b} Exciton transfer times were obtained from HEOM calculations, secular Redfield calculations and MLP predictions. Results are listed in Tab.~\ref{tab:transfer_times_bio_complexes}. For all four biological complexes, we find that the MLP predictions are more accurate than the secular Redfield calculations. 

		\begin{table}[!ht]
			\centering
			\begin{tabular}{lcccccc}
				\toprule
				\multirow{2}{*}{LHC} & \multicolumn{3}{c}{~~~~~Exciton transfer time [ps]~~~~~} & \multicolumn{3}{c}{~~~~~Computation time~~~~~} \\
					 & ~~~HEOM~ & ~MLP~ & ~Secular Redfield~ &   ~~~HEOM (GPU)~  & ~MLP (CPU)~ & Secular Redfield (CPU) \\
				\midrule
				FMO     &  5.78 &  4.99 &  4.18  & \unit[2.1]{min} & \unit[5]{ms}    & \unit[14.5]{min} \\
				RC      &  7.94 &  7.52 &  7.48  & \unit[3.9]{min} & \unit[3]{ms}    & \unit[8.9]{min}  \\
				CP43+RC & 13.80 & 11.64 & 11.16  & \unit[7.4]{h}   & \unit[5]{ms}    & \unit[14.6]{h}  \\
				CP47+RC & 18.92 & 19.31 & 15.08  & \unit[31.2]{h}  & \unit[4]{ms}    & \unit[40.9]{h}  \\
				\bottomrule
			\end{tabular}
			\caption{Comparison of exciton transfer times for the light-harvesting complexes (LHCs) considered in this study computed with the hierarchical equations of motion (HEOM), multi-layer perceptrons (MLPs) and secular Redfield. We report the obtained exciton transfer times as well as the runtimes of the calculations. Note, that HEOM calculations were run on GPUs, while MLP predictions and secular Redfield calculations were run on CPUs. MLPs were trained on PCA selected training sets and neither of the LHCs was included in the training sets. }
			\label{tab:transfer_times_bio_complexes}
		\end{table}

	\subsection{Computational cost of exciton dynamics calculations}\label{sec:transfer_time_results}

		Established methods for computing the population dynamics of excitonic systems such as the hierarchical equations of motion (HEOM) approach suffer from adverse computational scaling. Because of this drawback less sophisticated techniques with lower computational demand such as the secular Redfield method are more popular. Both methods need to run a full population dynamics calculation for obtaining exciton transfer properties such as the average transfer time or transfer efficiency. \\

		We report the runtimes of HEOM and secular Redfield calculations observed during the generation of the four datasets used in this study. Each dataset consists of 12000 randomly generated exciton Hamiltonians for which we computed average transfer times and transfer efficiencies with both methods (see Sec.~\ref{sec:dataset_generation} for details). HEOM calculations were carried out in the high-performance \textit{QMaster} package,\cite{Kreisbeck2011, Kreisbeck2012, Kreisbeck2014} which uses the architecture of GPUs for propagating a complete hierarchy of auxiliary matrices in parallel. Secular Redfield calculations were run on single core CPUs. Computation time for the individual Hamiltonians show some variations in computation time since depending on the excitation energy transfer times, we need to run the exciton dynamics for a larger or smaller number of time-steps. In Table~\ref{tab:transfer_time_computing_times} we show the average computation time for a single exciton Hamiltonian for each dataset (average over all 12,000 Hamiltonians), as well as the total computational cost to generate the complete datasets.\\

		As a compromise between accuracy and computational costs we truncate the hierarchy at $N_\text{max} = 5$ for the datasets with fewer pigments (RC and FMO) and at $N_\text{max} = 4$ for the datasets with more pigments (CP43 and CP47). We tested that the accuracy is retained at these truncation levels by simulating a random selection of 10 exciton Hamiltonians drawn from each dataset at respective lower and higher levels of truncation. We observe an average \unit[1.6]{\%} deviation in the calculated transfer times between the higher and lower truncation levels for the FMO dataset. The RC and CP43 datasets show smaller deviations with \unit[1.1]{\%} for RC and \unit[0.1]{\%} for CP43. For the CP47 we observe a deviation of about \unit[2.4]{\%}. 
		Despite the small deviations induced by the earlier truncation of the hierarchy, we use the lower truncation level in our dataset generation to keep the computational costs at a reasonable level (see Tab.~\ref{tab:transfer_time_computing_times}). \\

		While transfer properties in smaller systems, as they are modeled in the RC and FMO datasets, can be calculated within a few minutes, calculations on larger systems modeled in the CP43 and CP47 datasets show significantly higher computational demands with typical runtimes in the order of 6 to 18 hours. The significant computation times illustrate that large-scale simulations on artificial exciton systems can be computationally quite exhaustive. \\

		Multi-layer perceptrons (MLPs), however, encode a set of matrix operations, which allows for a significantly faster calculation of the properties of interest. By construction, the time for computing the output of an MLP scales linearly with the number of neurons in the architecture. We find that our trained MLP models could predict exciton transfer times and transfer efficiencies of one complete dataset introduced in this study in less than 10 seconds on a single CPU. This estimate includes the time spent for loading all 12000 exciton Hamiltonian matrices into memory and running the matrix operations encoded in the MLP architecture as well as writing the results of the prediction to file. \\

		\begin{table}[!ht]
			\centering
			\begin{tabular}{lcccccc}
				\toprule
				\multirow{2}{*}{Dataset~~~} & \multicolumn{2}{x{4cm}}{HEOM (GPU)~~} & \multicolumn{2}{x{4cm}}{MLPs (CPU)~~} & \multicolumn{2}{x{4cm}}{Secular Redfield (CPU)~~} \\
					& $T_\text{single}$  &   $T_\text{all}$      & $T_\text{single}$  &   $T_\text{all}$    & $T_\text{single}$  &   $T_\text{all}$ \\ 
				\midrule
				FMO  & \unit[4.0]{min} & \unit[33.6]{days}  & \unit[$<$0.1]{ms} & \unit[0.3]{s} & \unit[8.8]{min} & \unit[73.5]{days} \\
				RC   & \unit[2.2]{min} & \unit[18.3]{days}  & \unit[0.1]{ms} & \unit[1.6]{s} & \unit[4.4]{min} & \unit[36.3]{days} \\
				CP43 & \unit[5.8]{h}   & \unit[7.9]{years}  & \unit[0.6]{ms} & \unit[7.5]{s} & \unit[10.7]{h}  & \unit[14.7]{years} \\
				CP47 & \unit[18.2]{h}  & \unit[24.9]{years} & \unit[0.1]{ms} & \unit[1.5]{} & \unit[36.6]{h}  & \unit[41.9]{years} \\
				\bottomrule
			\end{tabular}
			\caption{Runtimes $T$ for exciton transfer time computations carried out with the \emph{QMaster} package, version 0.2,\cite{Kreisbeck2011, Kreisbeck2012, Kreisbeck2014} for all four datasets. We report runtimes for the entire datasets, $T_\text{all}$, each comprising of 12000 exciton Hamiltonians, as well as the average runtime per exciton Hamiltonian, $T_\text{single}$. Hierarchical equations of motion (HEOM) calculations were carried out on a NVIDIA Tesla K80 GPU and secular Redfield calculations as well as multi-layer perceptron (MLP) predictions on Intel(R) Xeon(R) CPUs X5650 @ \unit[2.67]{GHz}.}
			\label{tab:transfer_time_computing_times}
		\end{table}

		MLP models can be considered a valid alternative to secular Redfield calculations when running exciton dynamics calculations on a large number of excitonic systems due to the comparable accuracy at an orders of magnitude lower computational cost. \\

	\subsection{Dataset preparations for multi-layer perceptron predictions}\label{sec:mlp_comments}

		MLP models constructed in this study were designed to predict exciton transfer times and transfer efficiencies from the Frenkel exciton Hamiltonian of an excitonic system. The numerical values of excited state energies and inter-site couplings in the exciton Hamiltonian, as well as, the transfer times and transfer efficiencies depend on the chosen unit system. Further, the ranges of these properties can be very different (see for instance Tab.~\ref{tab:hamiltonian_ranges}). \\

		Features and targets with significantly different numerical values are more challenging to learn for most machine learning models as the applied model first has to learn the general range of numerical values before it can learn more subtle differences. The training procedure of machine learning models in general and MLPs, in particular, can be accelerated by rescaling features and targets to similar numerical values. \\

		Instead of providing the Frenkel exciton Hamiltonians as training features and the transfer times and efficiencies as training targets in a particular system of physical units we rescaled both features and targets based on the parameter ranges in the training set to facilitate faster and more stable MLP training. In particular, we subtracted the training set mean from all excited state energies in the exciton Hamiltonians and then mapped all feature elements $h_{ij}$ onto the interval $[-2, 2]$ (see Eq.~\ref{eq:feature_rescaling}), where $h_{ij}$ denotes a particular element of a particular exciton Hamiltonian and $H_\text{train}$ denotes the set of all exciton Hamiltonians in the training set. The rescaled features are denoted with $\tilde{h}_{ij}$. Rescaling the input features onto the $[-2, 2]$ interval ensures that all input values lie within sensitive regions of input layer activation functions, which avoids neuron saturation at the beginning of the MLP training.  \\

		\begin{align}\label{eq:feature_rescaling}
			\tilde{h}_{ij} = 4 \frac{h_{ij} - h_\text{min}}{h_\text{max} - h_\text{min}} - 2, \qquad\qquad h_\text{min} = \min\limits_{h_{ij} \in H_\text{train}}(h_{ij}), \qquad h_\text{max} = \max\limits_{h_{ij} \in H_\text{train}}(h_{ij}).
		\end{align}

		Prediction targets need to be rescaled onto an interval, which lies within the codomain of the output layer activation function. The target properties in this study are exciton transfer times and efficiencies. Both of these properties are always positive, but while efficiencies also have an upper bound, transfer times could a priori be arbitrarily large. We, therefore, chose the softplus function for the activation function of the MLP output layer as it exhibits similar properties. To use most of the non-linear regime of the softplus function, we decided to map our training targets $t$ (the collective set of transfer times and efficiencies) onto the interval $(0, 4]$ based on the maximum transfer time and efficiency in the training set $T_\text{train}$ (see Eq.~\ref{eq:target_rescaling}). Exciton transfer times and transfer efficiencies were rescaled separately. Note, that both exciton transfer times and efficiencies are always positive which justifies the implicit lower bound of zero in the equation. \\

		\begin{align}\label{eq:target_rescaling}
			\tilde{t} = \frac{4t}{t_\text{max}},\qquad\qquad t_\text{max} = \max\limits_{t \in T_\text{train}}(t)
		\end{align}

	\subsection{Bayesian optimization for hyperparameter selection}\label{sec:sup_bayesian_optimization}

		Multi-layer perceptrons (MLP) consist of a set of neurons organized in layers. Each neuron accepts an input, which is rescaled by a set of weights and biases intrinsic to the neuron, to calculate its output. The outputs of neurons in one layer are propagated through the MLP as the input for the subsequent layer. While weights and biases of each neuron are collectively referred to as the parameters of the MLP, an MLP model contains additional free parameters such as the number of layers and the number of neurons per layer. The latter are the hyperparameters of the model. MLP parameters are typically optimized on the training set with gradient based optimization techniques such as stochastic gradient descent. Hyperparameters are instead selected by optimizing the parameters of an MLP on the training set and evaluating the prediction accuracy on the validation set. \\

		In this study, we decided to employ a Bayesian optimization approach to find hyperparameters for accurate MLP architectures. We chose a total of six hyperparameters to be optimized and set fixed ranges for each of them for the Bayesian optimization. Hyperparameters and ranges are reported in Tab.~\ref{tab:bayes_opt_hyperparameter_selection}. In particular, we also included the number of training points as a hyperparameter to investigate by how much the prediction accuracy of MLPs can increase when expanding the training set. \\

		\begin{table}[!ht]
			\centering
			\begin{tabular}{lrr}
				\toprule
				Hyperparameter & low & high \\
				\midrule
				Training points   &       4000 &      6000 \\
				Network layers    &          3 &        18 \\
				Neurons per layer &          2 &      2000 \\
				Learning rate     &  $10^{-7}$ & $10^{-1}$ \\
				Regularization    & $10^{-18}$ &  $10^{6}$ \\
				\midrule
				\multirow{2}{*}{Activation function}  & \multicolumn{2}{c}{sigmoid, relu, tanh} \\
				                  & \multicolumn{2}{c}{softsign, softplus} \\
				\bottomrule
			\end{tabular}
			\caption{Selection of hyperparameters for Bayesian optimization of the MLP architecture. Lower and upper bounds were applied to the search space for five of the six parameters. Five different options were provided to the Bayesian optimizer for choosing an activation function for all but the last network layer.}
			\label{tab:bayes_opt_hyperparameter_selection}
		\end{table}

		The particular ranges for individual hyperparameters were chosen based on a few test training runs and chosen large enough that the Bayesian Optimizer is able to explore diverse MLP architectures. Although we found that MLPs perform more accurately when using more points in the training set, we restricted the number of training points to 6000 as a compromise between accuracy and computational demand. Especially for PCA sampled training sets we observed only a small advantage of larger training sets measured by the validation set error. Several activation functions were probed in the Bayesian optimization. \\


		\subsubsection{Convergence of Bayesian optimization runs}

		    Bayesian optimization selects a particular set of hyperparameters from all the possible values for each of the hyperparameters (see Tab.~\ref{tab:bayes_opt_hyperparameter_selection}). The MLP corresponding to this set of hyperparameters is then constructed and trained on the training set. Prediction errors on the validation set are used to evaluate the prediction accuracy of each constructed MLP after it was trained. \\

\begin{figure}[!ht]
			\centering
			\includegraphics[width = 0.6\columnwidth]{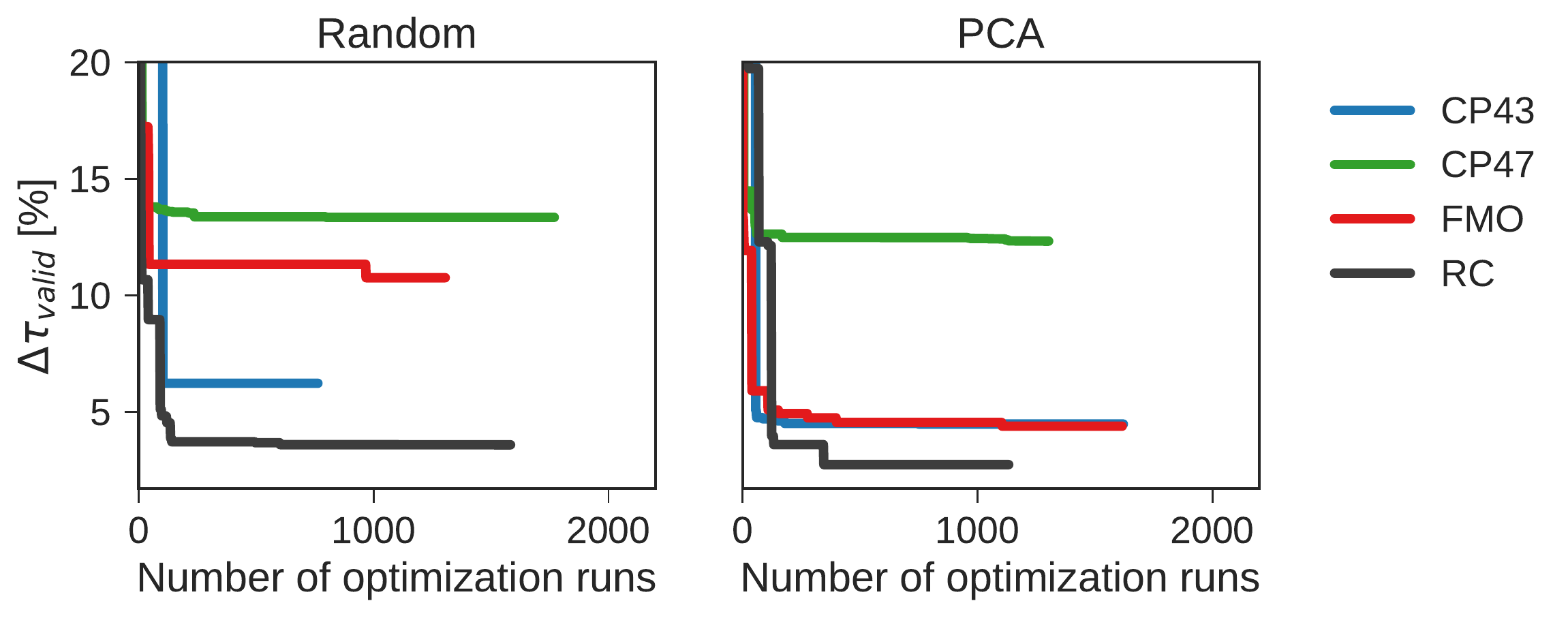}
			\caption{Smallest relative validation error for the set of MLPs trained during the Bayesian Optimization over the number of MLP evaluations in the optimization. 
			}
			\label{fig:bayesian_optimization_progress}
		    \end{figure}

		    The Bayesian optimization procedure was run for a total time period of seven days (walltime) on four GPUs (NVIDIA Tesla K80) for each dataset (FMO, RC, CP43, CP47) and each training set selection method (random, PCA). We extract the validation set error for all MLPs generated and trained in this process. During optimization, we keep track of validation error of the current optimal MLP architecture. Fig.~\ref{fig:bayesian_optimization_progress} to illustrate the progress of the hyperparameter optimization. After only a few iterations the prediction error for the validation set already has significantly dropped.

		    In each of the Bayesian optimization procedures we did not see any decrease in the validation set errors for at least the last 200 proposed MLP architectures. Further, we observe that as opposed to a single best set of hyperparameters the Bayesian optimization instead reveals a number of MLP architectures with different hyperparameter values but similarly small validation set errors. Therefore, we conclude that the Bayesian optimization converged for all datasets and identified reasonably accurate MLP architectures.

		\subsubsection{Bayesian optimization results}

\begin{figure}[!ht]
				\centering
				\begin{minipage}{0.45\textwidth}
					\flushleft
					A) FMO:\\
					\centering
					\includegraphics[width = 0.85\textwidth]{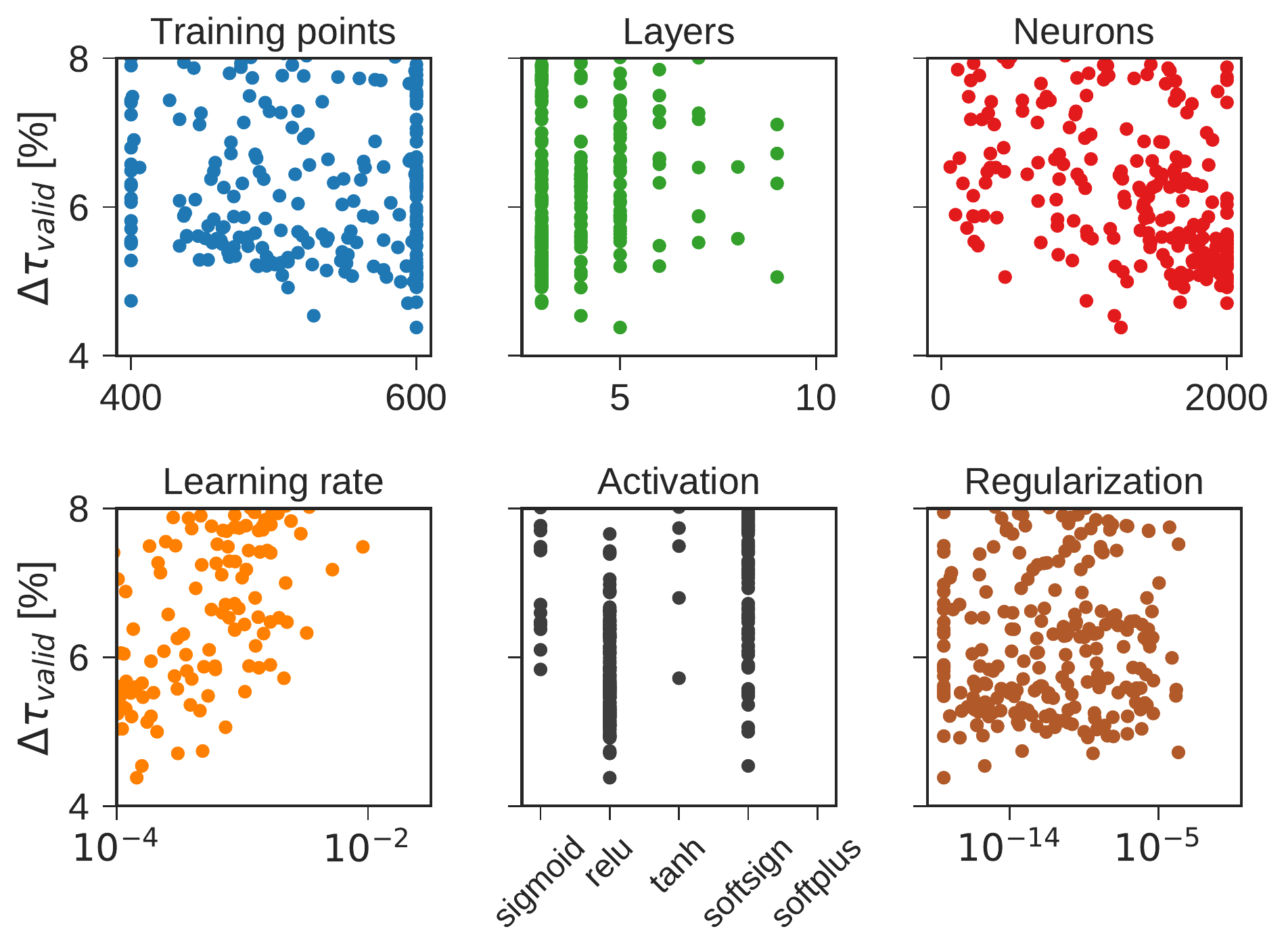} \\
				\end{minipage}
				\begin{minipage}{0.45\textwidth}
					\flushleft
					B) RC:\\
					\centering
					\includegraphics[width = 0.85\textwidth]{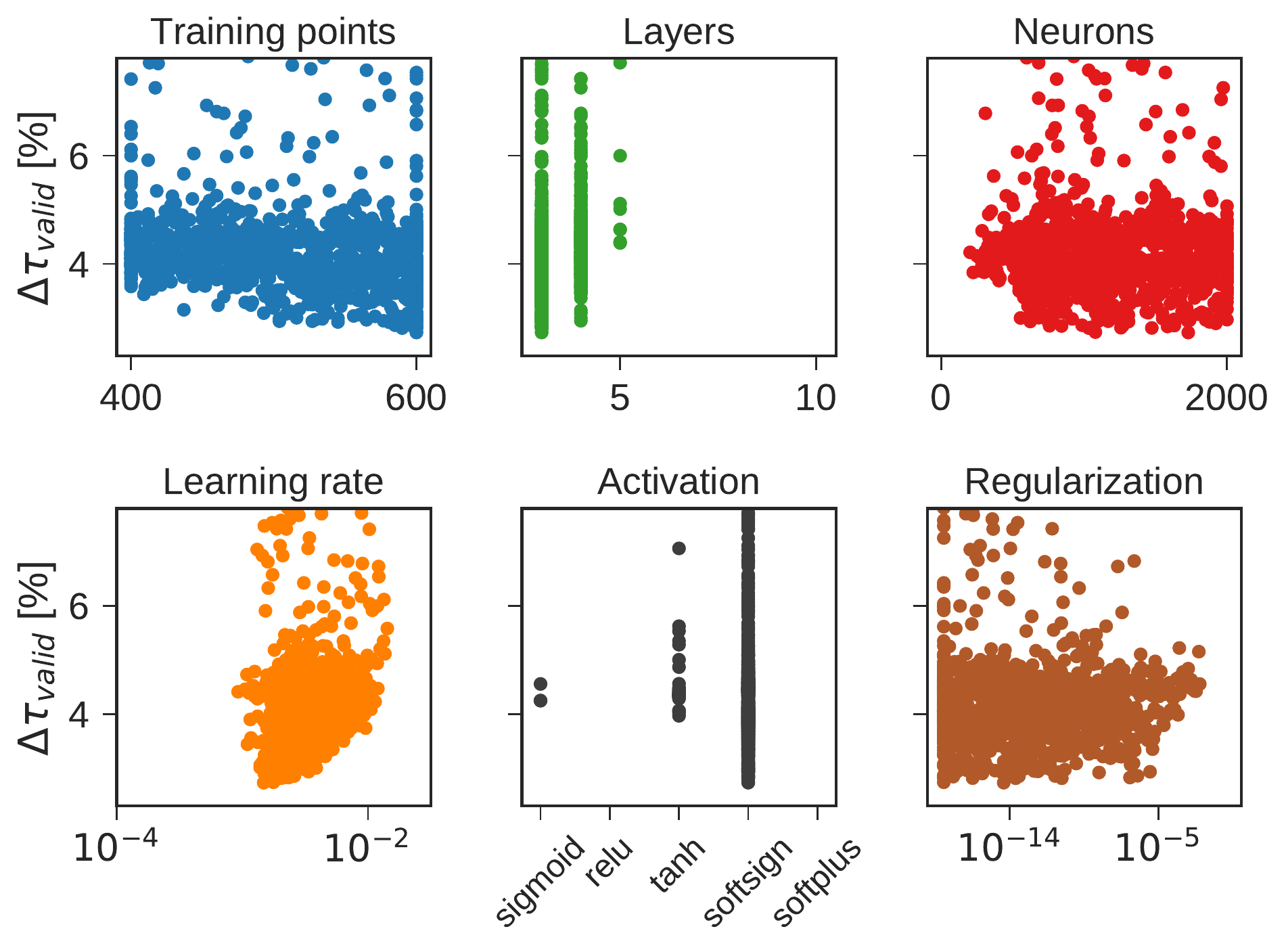} \\
				\end{minipage} \\
				
				\begin{minipage}{0.45\textwidth}
					\flushleft
					C) CP43:\\
					\centering
					\includegraphics[width = 0.85\textwidth]{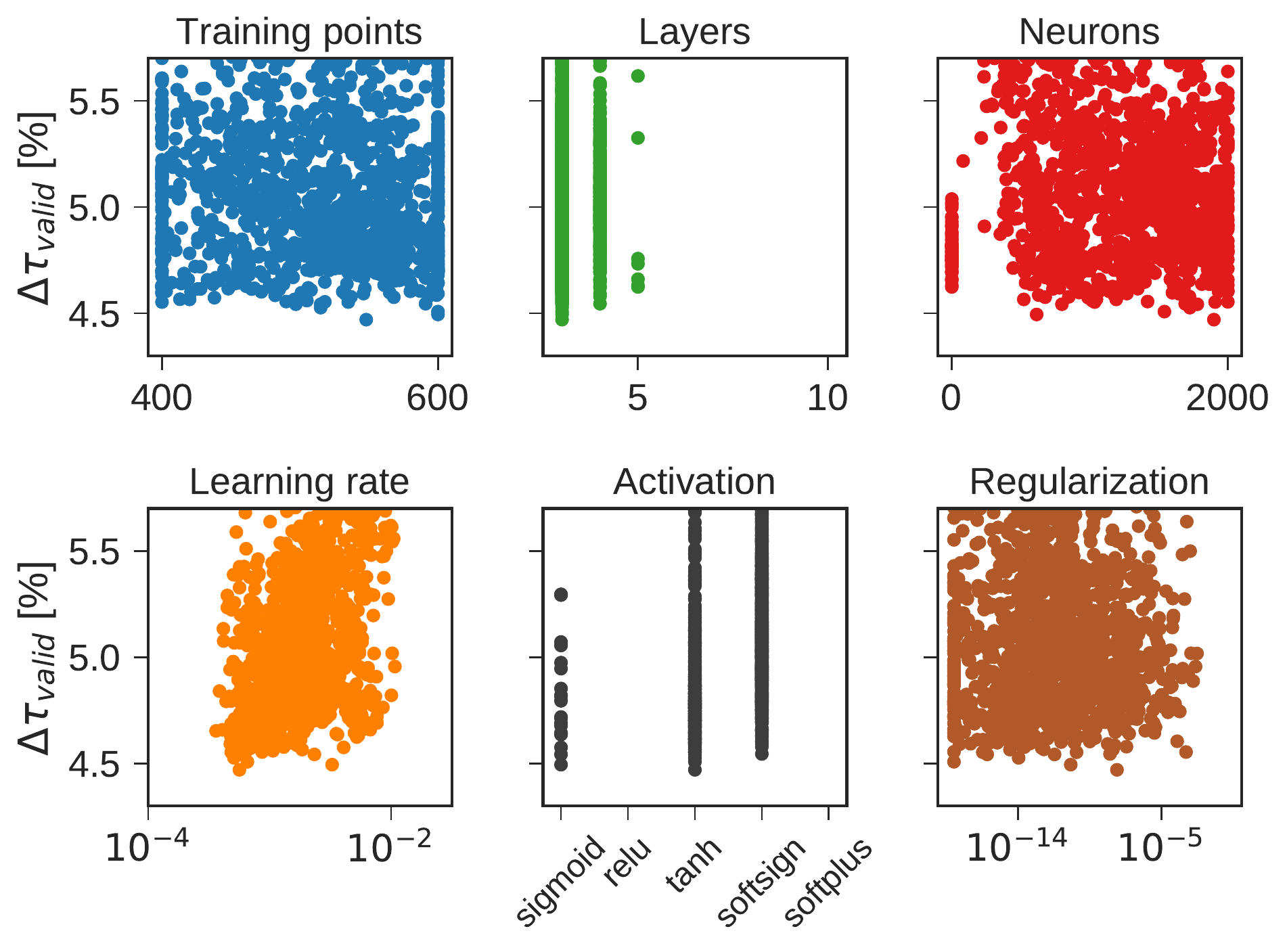}
				\end{minipage}
				\begin{minipage}{0.45\textwidth}
					\flushleft
					D) CP47:\\
					\centering
					\includegraphics[width = 0.85\textwidth]{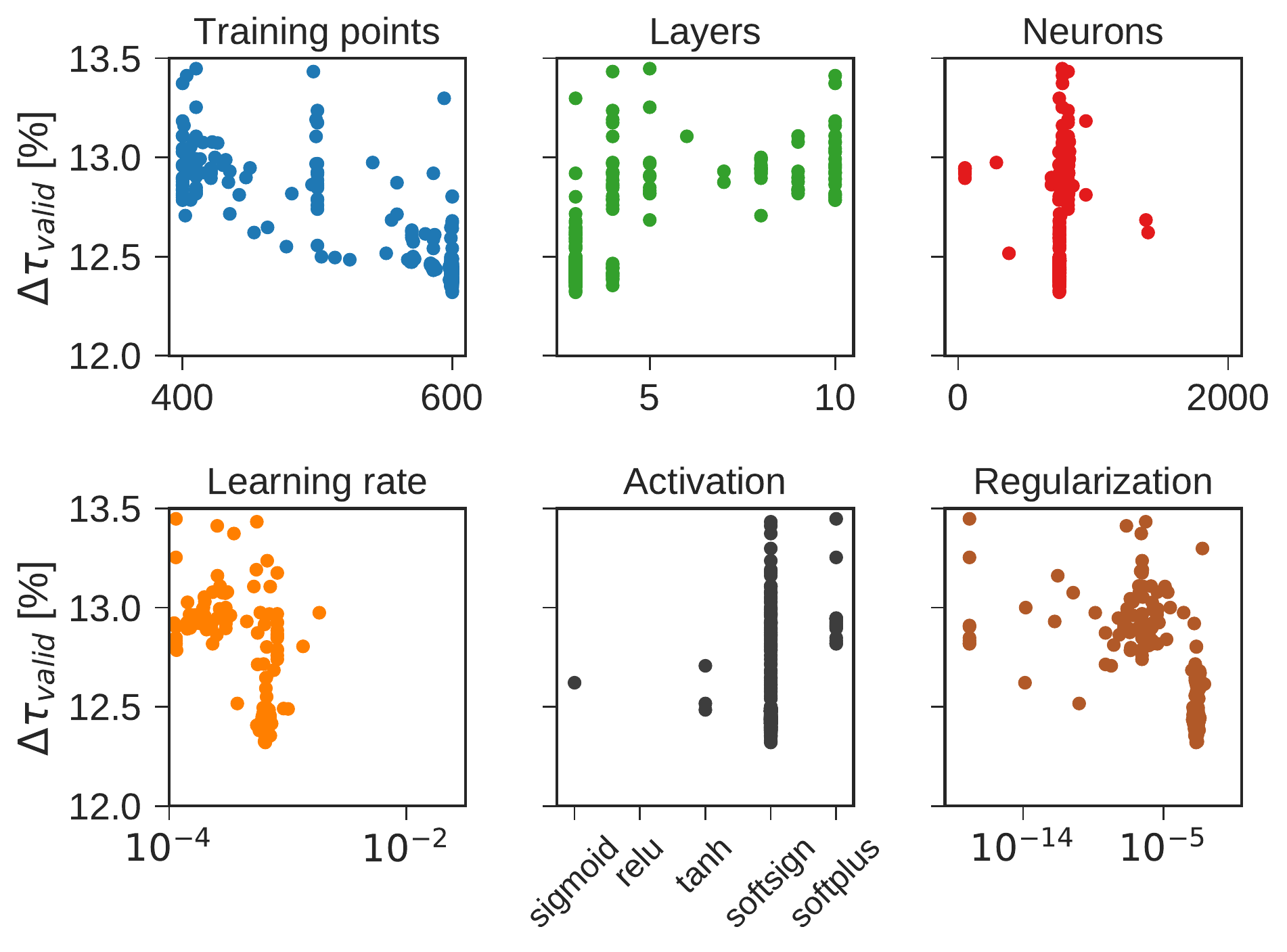}
				\end{minipage}
				\caption{Scatter plot of the average absolute relative errors of multi-layer perceptrons (MLPs) constructed during the Bayesian optimization procedure and trained on the indicated data set in dependence of the particular choices of hyperparameters made by the Bayesian optimizer. Each point represents the hyperparameters for the best architecture for each optimization step. MLPs were trained on PCA selected training sets.}
				\label{fig:remaining_bayes_opt_networks}
			\end{figure}

			We recorded the minimum validation set errors of all MLPs constructed and trained during the Bayesian optimization procedure to study the effect of particular hyperparameter choices on the prediction accuracy (see Fig.~\ref{fig:remaining_bayes_opt_networks}). 


			Optimal sets of hyperparameters resulting in the smallest validation set error for all four datasets with MLPs trained on PCA select (randomly drawn) training sets are reported in Tab.~\ref{tab:bayes_opted_hyperparameters} (Tab.~\ref{tab:bayes_opted_hyperparameters_random}). While MLPs trained on the RC and the FMO datasets, which consists of fewer excitonic sites, tend to prefer shallow but broad architectures we achieved the smallest relative validation set errors with more hidden layers and fewer neurons per hidden layer for the CP43 and the CP47 dataset with more excitonic sites. \\

			\begin{table}[!ht]
				\centering
				\begin{tabular}{lcccc}
					\toprule
					Hyperparameter     & RC            & FMO 		  & CP43 & CP47 \\
					\midrule 
					Training points    & 6000          & 6000         & 5480          & 6000          \\
					Layers             & 3             & 5            & 3             & 3             \\
					Neurons per layer  & 1729          & 1258         & 1899         & 749            \\
					Learning rate      & $10^{-3.83}$  & $10^{-3.44}$ & $10^{-3.24}$  & $10^{-3.19}$  \\
	 				Activation         & softsign      & relu         & tanh          & softsign      \\ 
					Regularization     & $10^{-14.37}$ & $10^{-18.0}$ & $10^{-7.80}$ & $10^{-2.91}$ \\
					\bottomrule
				\end{tabular}
				\caption{Optimized multi-layer perceptron (MLP) hyperparameters for the four investigated datasets (PCA selected training sets) used in the main text. Hyperparameters were optimized in a Bayesian optimization procedure. }
				\label{tab:bayes_opted_hyperparameters}
			\end{table}

			\begin{table}[!ht]
				\centering
				\begin{tabular}{lcccc}
					\toprule
						Hyperparameter    & RC & FMO & CP43 & CP47 \\
					\midrule
						Training points   &        6000  &  6000        &  6000        &  6000  \\
						Layers            &           3  &  3           &  3           &     7  \\
						Neurons per layer &         901  &  1382        &  1913        &   962  \\
						Learning rate     & $10^{-2.84}$ & $10^{-3.21}$ & $10^{-4.23}$ & $10^{-4.08}$ \\
						Activation        &     softsign &  softsign    & softsign     & softsign     \\
						Regularization    & $10^{-18.0}$ & $10^{-16.1}$ & $10^{-12.0}$ & $10^{-4.3}$ \\
					\bottomrule
				\end{tabular}
				\caption{Optimized multi-layer perceptron (MLP) hyperparameters for the four investigated datasets (randomly selected training sets). Hyperparameters were optimized in a Bayesian optimization procedure. }
				\label{tab:bayes_opted_hyperparameters_random}
			\end{table}

	\subsection{Comparison of exciton transfer times obtained from exciton dynamics calculations and multi-layer perceptron predictions}\label{sec:prediction_comparisons}

		We trained MLP models to predict exciton transfer times and efficiencies from Frenkel exciton Hamiltonians to provide an alternative to computationally costly exciton dynamics calculations. In this section, we comment on the prediction accuracy of MLP models by comparing their predictions with results obtained from two popular exciton dynamics approaches, the secular Redfield method and the hierarchical equations of motion (HEOM) formalism (on which the MLPs were trained). While secular Redfield is known to underestimate exciton transfer times it is computationally cheaper than HEOM. \\

\begin{figure}[!ht]
			\centering
			\includegraphics[width = 1.0\textwidth]{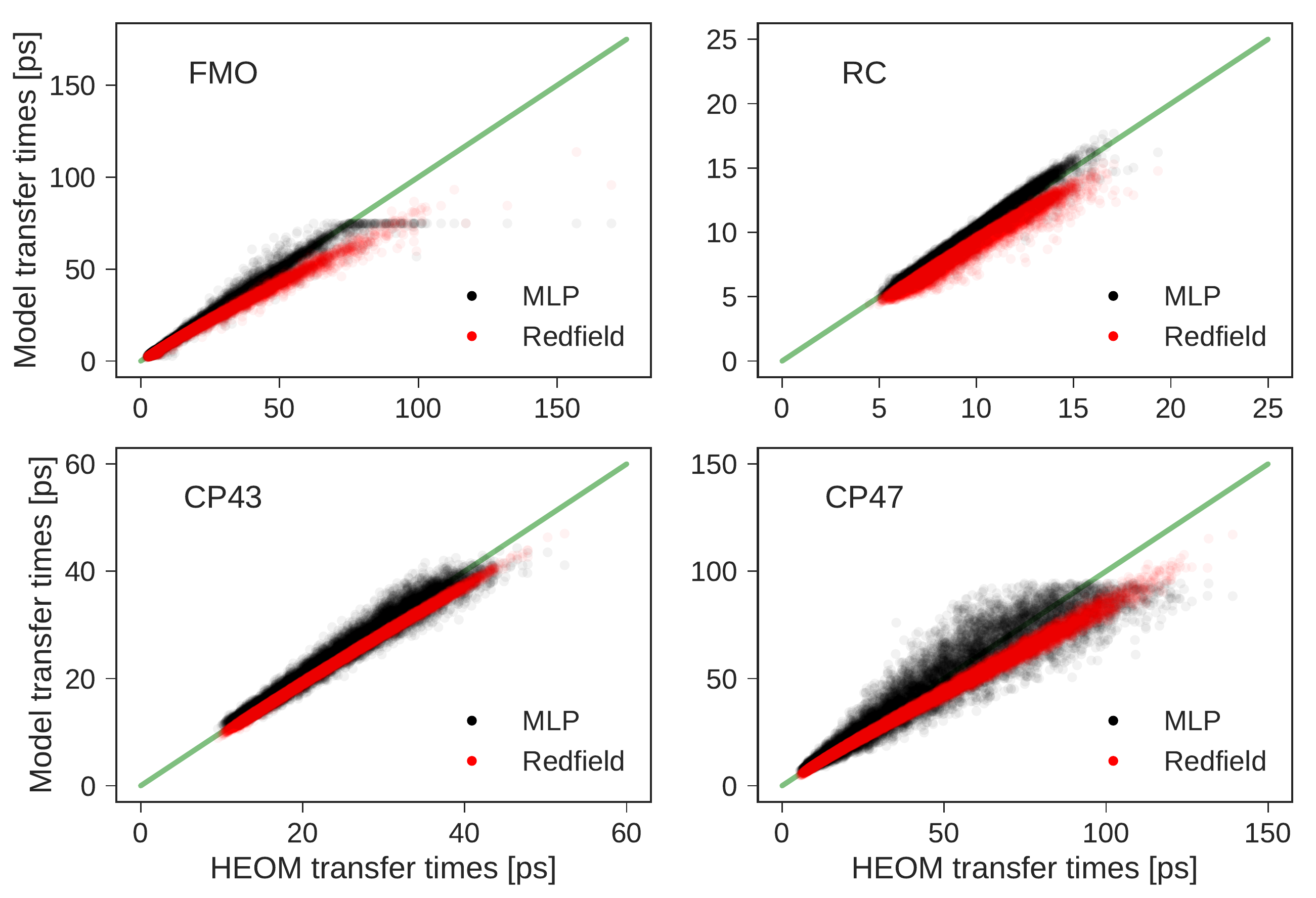}
			\caption{Exciton transfer times as computed with the hierarchical equations of motion (HEOM) approach compared to ecxiton transfer times calculated with the secular Redfield method or predicted from trained multi-layer perceptrons (MLPs). Panels show the exciton transfer times obtained for all 12000 exciton Hamiltonians in each of the four generated datasets. The griin line indicates perfect agreement between HEOM results and predictions by either MLPs or secular Redfield. }
			\label{fig:transfer_times_scatter_plot}
		\end{figure}

		We provide a context for MLP prediction accuracies by comparing MLP predicted transfer times to transfer times obtained from secular Redfield and HEOM calculations on the level of individual Hamiltonians. Fig.~\ref{fig:transfer_times_scatter_plot} shows scatter plots of transfer times obtained from all three approaches. We plot transfer times predicted by MLPs and secular Redfield versus transfer times calculated with HEOM, which we consider as the ground truth for the purpose of this study. The green line in Fig.~\ref{fig:transfer_times_scatter_plot} indicates perfect agreement between HEOM predictions and predictions by either MLPs or secular Redfield. \\

		By comparing the exciton transfer time predictions of the secular Redfield approach to the exciton transfer times of HEOM we observe that secular Redfield almost always underestimates transfer times consistently in all four datasets. MLPs instead over- and underestimate exciton transfer times, which is due to the symmetric loss function employed during MLP training and hyperparameter optimization. \\

		The absolute deviation between Redfield and HEOM transfer times generally increases with the value of the exciton transfer time, as observed in previous studies.\cite{Novoderezhkin2011} Also for the MLP predictions, we observe a larger deviation from the HEOM results for longer transfer times. However, in contrast to secular Redfield, the error is still distributed rather symmetrically around our ground truth. This observation is explained by the fact that MLPs were trained to minimize the relative deviation between predicted and HEOM transfer times. That is, for longer transfer times the predictions can afford larger deviations from HEOM which results in an opening funnel structure, especially seen in the scatter plot for the CP47 dataset. In addition, all of the presented datasets include fewer data points at larger transfer times (see Fig.~\ref{fig:transfer_time_distributions}), but MLPs are trained to minimize loss functions which take the unweighted average over all transfer times.




	\putbib[main]
	\end{bibunit}

\end{document}